# Students' Consistency of Graphical Vector Addition Method on 2-D Vector Addition Tasks

Jeffrey M. Hawkins*, John R. Thompson*[†], and Michael C. Wittmann*[†]

*Department of Physics and Astronomy, University of Maine, Orono, ME 04469, USA
[†]Center for Science and Mathematics Education Research, University of Maine, Orono, ME 04469, USA

**Abstract.** In a series of ten two-dimensional graphical vector addition questions with varying visual representations, most students stuck to a single solution method, be it correct or incorrect. Changes to the visual representation include placing vectors on a grid, making the vectors arrangements symmetric, varying the separation between vectors, and reversing the direction of either vector. We discuss the questions asked of students and their responses, emphasizing the results of one student who did change solution methods during an interview.

## INTRODUCTION

Vectors are used in many different physics topics from the introductory level through graduate classes. Research on student understanding of vectors shows that students often lack the ability to add vectors, which can lead to student difficulties.[1-5]

In our research, we have chosen to use different visual representations of graphical vector addition questions to investigate changes in student solution methods. There are many ways to ask students to add or subtract the same pair of vectors graphically. In addition to exploiting the translational invariance of vectors, various levels of representational detail can be provided. It is known that students' use many different methods to add vectors [5]; but the effects different visual representations have on which methods are used has not been studied.

To probe for such effects, we conducted a series of interviews consisting of graphical vector addition questions with varying representational detail. We chose interviews for the task because they would allow us to observe in some detail how students solved the vector addition questions.

## VECTOR ADDITION METHODS AND REPRESENTATIONS

Vectors to be added or subtracted can be presented in contact or separated; on or without a grid; at specific angles, such as parallel, relative to the grid; and so on. The more vectors one adds or subtracts, the richer the possibilities for varying representations. When choosing representations of vectors we chose to focus on 2-D vector addition with two vectors only. This choice greatly narrowed the possible arrangements we had to consider when developing our set of vector questions.

For each method of graphical vector addition we made a list of what representational changes we expected would either encourage or discourage their use. (See Table 1.) We considered several other methods we thought might be seen in the design of the interviews, but they were not used by any students in the interviews and are therefore not discussed here.

We considered students to be using the *head-to-tail* method when they arranged the vectors in a head-to-tail arrangement, connecting the free tail to the free head. For two vectors in a head-to-tail arrangement at a 90° angle, and parallel to the grid axes or without a grid, we considered students to be using *head-to-tail* if they redrew one of the vectors, or if they did not find the components of both of the vectors they were adding. Evaluating both vectors' components, then drawing the resultant, would be considered use of the *components* method described above. *Components* and *head-to-tail* are the only two methods discussed in the introductory text used in the course. [6,7]

We considered students to be using the *components* method when they split each vector into its $x$ and $y$

**TABLE 1. Graphical addition methods and representational changes which may encourage or discourage their use.**

| Method | Encourage Use | Discourage Use | Graphical Example of Method |
|---|---|---|---|
| Head-to-tail | Head-to-tail arrangement | Vectors are separated (not touching) <br> Vectors are on different grids <br> Non Head-to-tail arrangement | 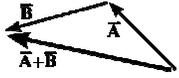 |
| Components | Grid <br> Only vertical or horizontal vectors <br> Vertical or horizontal symmetry | No grid | 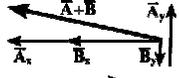 |
| Bisector | Tail-to-tail arrangement <br> Symmetric | Vectors are separated <br> Vectors are on different grids <br> Non Tail-to-tail arrangement | 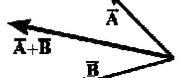 |
| General Direction | Head-to-head arrangement | Vectors are separated <br> Vectors are on different grids <br> Non Head-to-head arrangement | 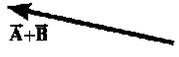 |

components individually and then added those components to find the components of the resultant vector. There are multiple methods students use to obtain the resultant vector from its components, especially when no grid is provided. In our categorization of this method, we did not consider the method they used to draw the resultant vector from the resultant $x$ and $y$ components.

We considered students to be using the *bisector* method if they arranged the vectors tail-to-tail and drew the resulting vector between them. Students who used the *bisector* method were notably sloppy about the length of the resultant vector.

We considered students to be using the *general direction* method when they drew a vector with the general magnitude and direction they thought the resultant vector would have without redrawing any vectors or using the components of the vectors to find the resultant vector.

## INTERVIEWS

The interviews consisted of ten vector questions, five with grids and five without grids, which had various arrangements of vectors as described above. The interview questions were ordered so that five vector questions (all either with a grid or without one) were followed by some other distracter task, and then the other five vector questions. We anticipated the presence of the grid to be the variable that would have the most impact on the student responses, so it was the variable that was changed before and after the distracter task. The distracter task was either on sound waves or on integration and usually took about half an hour. The intent of adding a distracter question was to cause the students to approach the second set of vector tasks without definite memory of the methods used in the first set.

Subjects were interviewed either late in the first semester of the introductory calculus-based course or early in the second semester of the introductory algebra-based course. The main goal of the interviews was to check for consistency rather than checking for correct understanding. The skill level of the students ranged from students who answered all the questions correctly to students who answered all the questions incorrectly. We found that most students consistently used the same general solution method even when the representation of the vectors changed.

## Strong Evidence for Basic Understanding

All of the students in the interviews could translate vectors without trouble. All students could give textbook definitions of vectors as something with a magnitude or length and a direction. Some students even expressed explicitly that two vectors are the same as long as the directions and magnitudes were the same, and that the location of the vector was not important. Most students expressed relief when they were given questions with a grid.

## Consistency of Method Use

Of the 8 students interviewed, 7 students used only one method of graphical vector addition during the interview. (See Table 2.) The only student to change methods was Student H, and his interview will be discussed later. The other students used one of three methods.

Three students (A, B and C) used the *head-to-tail* method to answer all 10 questions. When the grid was added, the students all used the grid to more accurately redraw the vectors into a head-to-tail arrangement before drawing the resulting vector.

Students D and E both used the *components* method to complete all the questions even though they received the questions in a different order: Student E was the only student to receive the non-gridded questions first and the gridded questions second. The order within the two sets of five questions was maintained, but the order of the five questions sets was flipped for this student.

On the questions without a grid, Student D drew the components for each vector on an axis he made up, separately added the *y* and *x* components of each vector, and used the resultant *x* and *y* vectors to get the resultant vector. When Student D was given the questions with a grid, he used the grid to find the components along the given axis, and then added them together to find the resultant vector.

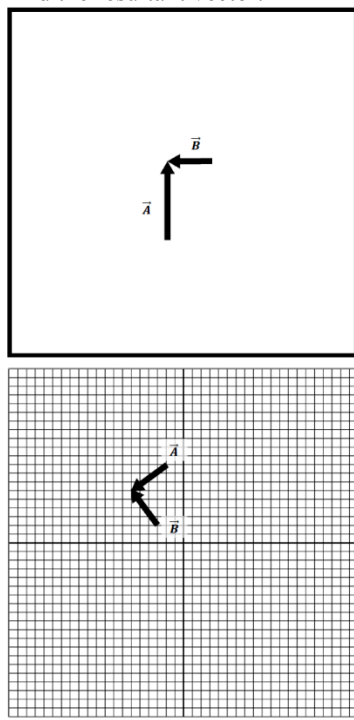

**FIGURE 1.** Question 5 (top) and Question 8 (bottom). Students were asked to "Add the two vectors $\vec{A}$ and $\vec{B}$ below to get a new vector $\vec{R}$ where $\vec{R} = \vec{A} + \vec{B}$."

**TABLE 2.** Student solution methods by student.

| Student | Method |
|---|---|
| Student A | Head-to-Tail |
| Student B | Head-to-Tail |
| Student C | Head-to-Tail |
| Student D | Components |
| Student E | Components |
| Student F | Bisector |
| Student G | General Direction |
| Student H | Various |

On the gridded questions, Student E used the grid to find the components and add them to get the resultant vector the same way Student D did. On the questions without a grid, Student E sketched in his own grid lines over each vector at the same scale and orientation as before. Student E then followed the same procedure he used on the gridded questions to get numerical components, and drew a separate grid to draw the resultant vector.

These two students did not do what we expected students to do when they answered these questions. Neither student changed from the *components* method when an obstruction to its use was present in the question. Instead they persisted, one student starting with the method despite not having a grid, and the other continuing to use the method despite not having a grid.

Student F used the bisector method to answer all 10 questions. Student F first drew the vectors tail-to-tail and then drew a vector between them, stating that the resultant vector's direction and length was determined by the lengths of the two vectors. Student F was very consistent with the use of this method, but did question it briefly. On Question 8 (gridded; see Fig.1) Student F mentioned that he thought maybe the vector should be longer than he had drawn it, based on using *component* reasoning. But he decided it was too simple and stuck with the bisector method.

Student G used a mix of both the *general direction* method and the *bisector* method. He stated that you had two vectors, and that the resultant vector would be in a direction between the two vectors. However, he simply drew the resultant vector rather than redrawing either vector to create a tail-to-tail arrangement. This led us to think he could have been using the *bisector* method based on some of his description; however, he could have been using the *general direction* method based on his drawings. Student G expressed the most discomfort with not having a grid behind the vectors. When a grid was present, he tried to change methods: he wrote down the slopes of the vectors, looked at them for awhile, decided he did not know how to use the components of the vectors to add them together, and went back to his initial method. This consideration of another method was more than most others did during their interviews.

## Observing a Change of Methods

The only student who changed methods at all was Student H, one of the students in the calculus-based course. The first change in methods Student H made was a switch from the *bisector* method to the *head-to-tail* method on question 5 (Fig. 1). Question 5 is a head-to-head arrangement of vectors at a 90° angle with both vectors parallel to the walls of the box they were enclosed in. The student had previously seen vectors that were parallel to the enclosing walls but separated, and vectors that were not separated, but also not parallel to the enclosing walls. We present interview excerpts to describe the transitions in methods used.

Student H: *"Ok."* (Long pause) *"I can't remember if you use the head-to-tail method. Hold on, I'm going to run through the head-to-tail, just… I think that's for subtraction… if there is such a thing."* (while drawing vectors head-to-tail, then drawing the resultant vector)

After Student H has finished drawing the resultant:

Student H: *"So yeah, that makes sense, cause if you... Yeah, head-to-tail does basically the same thing that I would do if I drew the vector down here* (draws vectors tail-to-tail, but does not draw bisector)*, but, you know, anyway, so there's R* [the resultant vector]*."*

Between Questions 5 and 6, the student answered the distracter task. When starting the second set of vector questions, Student H decided to continue using the *head-to-tail* method. The student explained, *"I'm not doing it the whole eyeballing way that I did at the beginning because, in doing these problems I've realized that this* [the head-to-tail method] *is probably a more sound way, considering that we actually went over it in class."*

The student's stated reason for switching was because of recall of the classroom method. Student H used one other method, as well. On question 8, Student H mentioned that he expected the resultant vector to "*be like up one square*" using components reasoning, before using the *head-to-tail* method: *"So now… we're adding…* (pointing at A) *one, two, three… (pointing at B) this is, one, two, three, four. Well, ok… Yeah, ok. The same, the same way is gonna work. I was thinking about it because, like, I thought that the y since this one* (pointing at A) *point goes down three squares but this one* (pointing at B) *goes up four squares. Kinda looks like that, one, two, three, four. Then the resultant vector is gonna be like up one square, I guess."*

Interviewer interrupts with *"Ok."*

Student H continues *"Cause they kind of cancel each other out, but I was just making sure that if I use that head-to-tail method again, then it will work."*

In this dialogue the student is clearly using the *components* method to describe the addition of the vertical components of the two vectors, but the student does not comment on the horizontal components of any of the vectors. The student does not use components reasoning to complete the addition of the problem, and returns to the *head-to-tail* method to do so. The student does discuss the agreement of the addition of the vertical components and the addition by the *head-to-tail* method: *"And… as I… assumed, you know, it does actually travel up one, one square,* (interviewer interrupts with "ok") *so. Kind of makes me think I was doing it right."*

After this brief use of *components* reasoning, the student does not comment on the components of the vectors in the following questions and continues to use the *head-to-tail* method exclusively to add the vectors.

## CONCLUSIONS

Interviews using very basic 2-D vector addition questions were designed to trigger different solution methods in students, based on representational details. We observed that most students stuck to one method for the entire interview despite changes in details that may change the relative favorability of that method. This suggests that in most cases either students lack a rich toolbox for solving vector addition problems, that they do not recognize the utility of multiple methods when solving problems, or that the tasks weren't as strongly cuing as we expected. Our preliminary results indicate that it would be beneficial to develop more refined ways of understanding if and when student responses change due to contextual and representational cues, and which cues or cue combinations affect responses.

Finally, Question 8 prompted two students to talk about the components of the individual and resultant vectors. Student F only discussed the horizontal components; Student H only discussed the vertical components. Question 8 was the first question presented to these students that had both a head-to-head arrangement of vectors and a grid. There may be something about the cues or cue combinations in Question 8 that causes students to consider changing methods on this particular question that should be investigated further.

## ACKNOWLEDGMENTS

This work was supported in part by NSF Grant #DRL-0633951.

## REFERENCES

1. R. D. Knight, *Phys. Teach.* **33**, 74-78 (1995).
2. N.-L. Nguyen and D. E. Meltzer, *Am. J. Phys.* **71**, 630-638 (2003).
3. S. Flores, S. E. Kanim, and C. H. Kautz, *Am. J. Phys.* **72**, 460- 468 (2004).
4. P. S. Shaffer and L. C. McDermott, *Am. J. Phys.* **73**, 921-931 (2005).
5. J. Van Deventer, "Comparing student performance on isomorphic math and physics vector representations," Masters Thesis, The University of Maine, 2008.
6. J. S. Walker, *Physics Third Edition*, Upper Saddle River, NJ: Prentice Hall, 2007, pp. 61-64.
7. R. D. Knight, *Physics for Scientists and Engineers Second Edition: A Strategic Approach*, Reading, MA: Addison Wesley, 2008, pp. 75-86.